\documentclass[12pt]{article}
\usepackage{mathrsfs}
\usepackage{amsfonts}
\usepackage{latexsym,amssymb,amsmath,amscd,amsfonts,epsfig,psfrag,graphicx}
\usepackage{lineno}
\newtheorem{thm}{Theorem}[section]

\newtheorem{lem}[thm]{Lemma}
\newtheorem{alg}[thm]{Algorithm}
\def\pf{\noindent{\it Proof.\;\;\:}}
\def\qed{\nopagebreak\hfill{\rule{4pt}{7pt}}
\medbreak}

\title{\bf Dynamic 3-Coloring of Claw-free Graphs
\footnote{Supported by NSFC, PCSIRT and the ``973" program.}}

\author{
\small Xueliang Li and Wenli Zhou\\
\small Center for Combinatorics and LPMC-TJKLC, Nankai University\\
\small Tianjin 300071, China. Email: lxl@nankai.edu.cn, louis@cfc.nankai.edu.cn\\
}
\date{}
\begin{document}

\maketitle

\begin{abstract}
A {\it dynamic $k$-coloring} of a graph $G$ is a proper $k$-coloring
of the vertices of $G$ such that every vertex of degree at least 2
in $G$ will be adjacent to vertices with at least 2 different
colors. The smallest number $k$ for which a graph $G$ can have a
dynamic $k$-coloring is the {\it dynamic chromatic number}, denoted
by $\chi_d(G)$. In this paper, we investigate the dynamic
3-colorings of claw-free graphs. First, we prove that it is
$NP$-complete to determine if a claw-free graph with maximum degree
3 is dynamically 3-colorable. Second, by forbidding a kind of
subgraphs, we find a reasonable subclass of claw-free graphs with
maximum degree 3, for which the dynamically 3-colorable problem can
be solved in linear time. Third, we give a linear time algorithm to
recognize this subclass of graphs, and a linear time algorithm to
determine whether it is dynamically 3-colorable. We also give a
linear time algorithm to color the graphs in the subclass by 3
colors.
 \\[2mm]

\end{abstract}

\section{Introduction}

We follow the terminology and notations of \cite{Bondy} and, without
loss of generality, consider simple connected graphs only.
$\delta(G)$ and $\Delta(G)$ denote, respectively, the minimum and
maximum degree of a graph $G$. For a vertex $v\in V(G)$, the {\it
neighborhood} of $v$ in $G$ is $N_G(v)=\{u\in V(G): u$ is adjacent
to $v$ in $G\}$, and the degree of $v$ is $d(v)=|N_G(v)|$. Vertices
in $N_G(v)$ are called {\it neighbors of} $v$. $P_n$ denotes the
path on $n$ vertices. A subset $S$ of $V$ is called an {\it
independent set} of $G$ if no two vertices of $S$ are adjacent in
$G$. An independent set $S$ is {\it maximum} if $G$ has no
independent set $S'$ with $|S'|>|S|$. The number of vertices in a
maximum independent set of $G$ is called the {\it independence
number} of $G$ and is denoted by $\alpha(G)$.

For an integer $k>0$. A {\it proper $k$-coloring} of a graph $G$ is
a surjective mapping $c:\ V(G)\rightarrow\{1,2,\ldots,k\}$ such that
if $u, v$ are adjacent vertices in $G$, then $c(u)\neq c(v)$. The
smallest $k$ such that $G$ has a proper $k$-coloring is the {\it
chromatic number} of $G$, denoted by $\chi (G)$.

The {\it dynamic coloring} of a graph $G$ is defined as a proper
coloring of $G$ such that any vertex of degree at least 2 in $G$ is
adjacent to more than one color class. For an integer $k>0$, a {\it
proper dynamic $k$-coloring} of a graph $G$ is thus a surjective
mapping $c: V(G)\rightarrow \{1,2,\ldots,k\}$ such that both of the
following two conditions hold:
\begin{itemize}
\item[(C1)] if $u,v \in V(G)$ are adjacent vertices in $G$, then
$c(u)\neq c(v)$; and \item[(C2)] for any $v\in V(G)$,
$|c(N_G(v))|\geq min\{d(v), 2\}$, where and in what follows,
$c(S)=\{c(u)|u\in S$ for a set $S\subseteq V(G)\}$.
\end{itemize}

We call the first condition, which characters proper coloring, the
{\it adjacency condition}, and we call the second condition the {\it
double-adjacency condition}. The smallest integer $k>0$ such that
$G$ has a proper dynamic $k$-coloring is the {\it dynamic chromatic
number} of $G$, denoted by $\chi_d (G)$.

In order to show the results in this paper, we will give some new
definitions. Similar to the definition of the dynamic coloring, a
{\it dynamic $k$-edge-coloring} of a graph $G$ is a proper
$k$-edge-coloring of $G$ such that every edge with at least 2
adjacent edges in $G$ will be adjacent to edges with at least two
different colors. The smallest number $k$ for which a graph $G$ can
have a dynamic $k$-edge-coloring is the {\it dynamic edge chromatic
number}, denoted by ${\chi_d}'(G)$.

The dynamic chromatic number has very different behaviors from the
traditional chromatic number. For example, from \cite{Lai 3} we know
that for many graphs $G$, $\chi_d(G-v)>\chi_d(G)$ for at least one
vertex $v$ of $G$, and there are graphs $G$ for which
$\chi_d(G)-\chi(G)$ may be very large.

From \cite{Lai 2} we know that if $\Delta (G)\leq 2$, we can easily
have a polynomial time algorithm to give the graph $G$ a dynamic
$\chi_d$-coloring. In \cite{Lai 1}, Lai, Montgomery and Poon got an
upper bound of $\chi_d(G)$ that if $\Delta(G)\geq 3$, then
$\chi_2(G)\leq \Delta(G)+1$. The proof is very long compared with
the proof of a similar result in the traditional coloring. In
\cite{Lai 2} and \cite{Lai 3}, Lai, Lin, Montgomery, Shui and Fan
got many new and interesting results on the dynamic coloring.
Recently, in \cite{Li 1} we proved that it is $NP$-complete to
determine if a triangle-free graph with maximum degree 3 is
dynamically 3-colorable. This is a little interesting because we
know that for graphs $G$ with $\Delta(G) = 3$ the $3$-colorable
problem of the traditional vertex coloring can be solved in
polynomial time.

Let $G$ be a graph with maximum degree 3. We define a family of
subgraphs $A_i$ of $G$, in which every $A_i$ is a path with $i$
vertices such that the $i-2$ internal vertices have degree 2 in $G$,
and the two end-vertices have degree 3 in $G$. A {\it pendant path}
of a graph $G$ is such a path that the internal vertices have degree
2 in $G$, one end-vertex has degree 1 and the other end-vertex has
degree 3. In the present paper, we concentrate on the dynamically
3-colorable problem for claw-free graphs. First, we prove that for a
claw-free graph $G$ with $\Delta (G)=3$ it is still $NP$-complete to
decide if $G$ is dynamically 3-colorable. This is also an
interesting result which is different from a result for the
traditional colorings. In order to find some kind of graphs for
which the dynamically 3-colorable problem is polynomially solvable,
we consider the subclass of the claw-free graphs with maximum degree
3, in which every graph is $A_i$-free $(i=3j+1, \ j\in
\mathbb{Z}^{+})$. We find that this kind of graphs can be recognized
in $O(n)$ time, and it can be done in $O(n)$ time to determine
whether they are dynamically 3-colorable, and we will give an $O(n)$
time algorithm to find a dynamic 3-coloring of the graphs.

\section{$NP$-complete results}

In \cite{Lai 2}, the authors proved the following theorem:

\begin{thm}\label{thm1}
If $G$ is claw-free, then $\chi_d(G)\leq \chi(G)+2$, and the
equality holds if and only if $G$ is a cycle of length 5 or of
even length not a multiple of 3.
\end{thm}

So, apart from some special cycles, the difference between the
dynamic chromatic number and the chromatic number for claw-free
graphs is at most one. If we know $\chi(G)$ and there would be a
polynomial time algorithm to determine $\chi(G)=\chi_d(G)$ or
$\chi(G)=\chi_d(G)+1$ except the special cycles described in Theorem
\ref{thm1}, we can get some results on dynamic colorings by those on
the traditional colorings for claw-free graphs. But unfortunately,
we will show that even if we know the chromatic number of claw-free
graphs, we cannot get the dynamic chromatic number in polynomial
time unless $P=NP$. By the relation between the edge-coloring of a
graph $G$ and the vertex coloring of the line graph $L(G)$ of $G$,
we will get the result immediately after we finish the proof of the
following Theorem \ref{thm2}.

First, we give a formal definition of the {\it dynamically
3-edge-colorable problem}, denoted by {\bf Dy-3-Edge-Col}, which is
stated as follows:

{\bf Input}: A bipartite graph $B=B(V,E)$ and $\Delta(B)=3$.

{\bf Question}: Can one assign each edge a color, so that only 3
colors are used and this is a dynamic edge-coloring ? i.e., is
${\chi_d}'(B)\leq 3$ ?

In \cite{Holyer}, the author proved that it is $NP$-complete to
determine whether a cubic graph is 3-edge-colorable. We will use the
result to prove that the Dy-3-Edge-Col is $NP$-complete.
\begin{thm}\label{thm2}
The Dy-3-Edge-Col is $NP$-complete.
\end{thm}
\pf First, it is obvious that the problem is in $NP$.

\begin{figure}[h]
\begin{center}
\includegraphics[width=6cm]{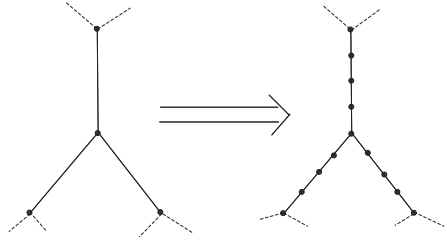}
\caption{The local transformation}\label{eps1}
\end{center}
\end{figure}

Second, given a cubic graph $C$. For every edge in $C$, we will use
a $P_5$ to replace the edge and construct a new graph $B$, i.e., we
subdivide every edge exact 3 times. The local transformation is
shown in Figure \ref{eps1}.

It is easy to see that $C$ is 3-edge-colorable if and only if $B$ is
dynamically 3-edge-colorable. And, by the structure of $B$, the
length of every cycle of $B$ is a multiple of 4. So $B$ does not
have any odd cycles, and thus is a bipartite graph. It is obvious
that $\Delta(B)=3$. Since the 3-edge-colorability for cubic graphs
is $NP$-complete, the Dy-3-Edge-Col must be $NP$-complete. \qed

\noindent {\bf Remark.} In the proof, we can subdivide each edge
$3j$ times for some $j\in Z^+$, instead of 3 times, and the proof
can still hold. Different edge could use different $j$. Therefore,
we have the following stronger statement.
\begin{thm}\label{thm2'}
It is $NP$-complete to determine whether a graph $G$ is dynamically
3-edge-colorable, obtained from a cubic graph $C$ by subdividing
each edge of $C$ $3j$ times for some $j\in Z^+$.
\end{thm}

For traditional edge-colorings, if a graph $G$ is bipartite, then
$\chi'(G)=\Delta(G)$ and there is a polynomial time algorithm to
color it. So, Theorem \ref{thm2} is different from the result for
traditional edge-colorings.

Next, we give a formal definition of {\it dynamically 3-colorable
problem}, denoted by {\bf Dy-3-Col}, which is stated as follows:

{\bf Input}: A graph $G=G(V,E)$.

{\bf Question}: Can one assign each vertex a color, so that only 3
colors are used and this is a dynamic coloring ? i.e., is
${\chi_d}(G)\leq 3$ ?

By the structure of the bipartite graph $B$ in the proof of Theorem
\ref{thm2}, we know that $L(B)$ is a line graph with maximum degree
3. Notice that a graph $G$ is dynamically $k$-edge-colorable if and
only if the line graph $L(G)$ of $G$ is dynamically $k$-colorable.
So, we have
\begin{thm}\label{thm3'}
It is $NP$-complete to determine whether the line graph $L(B)$ with
maximum degree 3 is dynamically 3-colorable. As a result, it is
$NP$-complete to determine whether a line graph with maximum degree
3 is dynamically 3-colorable.
\end{thm}
Since line graphs are claw-free graphs, then we have
\begin{thm}\label{thm3}
For claw-free graphs $G$ with $\Delta(G)=3$, the Dy-3-Col is
$NP$-complete.
\end{thm}

For traditional colorings, it is polynomially solvable whether a
graph $G$ is 3-colorable when $\Delta(G)\leq 3$. So we can see that
the dynamic coloring problem is very difficult to deal with even for
claw-free graphs with maximum degree 3. In next section we will find
some reasonable kind of graphs in which we can determine if a graph
is dynamically 3-colorable in polynomial time. Theorems \ref{thm2'}
and \ref{thm3'} could be omitted as intermediate results. But, we
prefer to list them in order to understand why we choose to study
this kind of graphs in next section.

\section{A polynomial time result}

From Theorem \ref{thm3}, the Dy-3-Col is $NP$-complete for claw-free
graphs with maximum degree 3, and because of Theorem \ref{thm3'},
the problem is $NP$-complete even for the line graph $L(B)$, where
$B$ is built up from a cubic graph by subdividing every edge exact 3
times. By reviewing the proof of Theorem \ref{thm2}, we notice that
there are many $A_4$ in $L(B)$. The question is: can the Dy-3-Col be
solved in polynomial time for both claw-free and $A_4$-free graphs
$G$ with $\Delta(G)=3$ ? The answer is No, because in the local
transformation, we can use any $P_i$ $(i=3j+2, \ j\in
\mathbb{Z}^{+})$ to replace the edges of the cubic graph $C$ to get
another graph $B'$, and $C$ is 3-edge-colorable if and only if $B'$
is dynamically 3-edge-colorable. Although $B'$ may not be bipartite,
we can still get Theorem \ref{thm3'}. So, another question is: can
the Dy-3-Col be solved in polynomial time for both claw-free and
$A_i$-free (for all $i=3j+1, \ j\in \mathbb{Z}^{+}$) graphs $G$ with
$\Delta(G)=3$ ? The answer is Yes. For convenience, we denote by
$\mathscr{C}$ the set of graphs $G$ with $\Delta(G)\leq3$ which are
both claw-free and $A_i$-free $(i=3j+1, \ j\in \mathbb{Z}^{+})$.
Then we have
\begin{thm}\label{thm4}
The Dy-3-Col is polynomially solvable for graphs in $\mathscr{C}$.
\end{thm}
\pf Given a graph $G$ in $\mathscr{C}$. First, delete all the
vertices in the pendant paths of $G$ except the end-vertices of
degree 3, to get the first graph $G_1$. It is easy to see that $G$
is dynamically 3-colorable if and only if $G_1$ is dynamically
3-colorable. Then $G_1$ has vertices of only degrees 2 and 3.
Second, delete all the internal vertices in $A_i$ $(i=3j+2, \ j\in
\mathbb{Z}^{+})$ of $G_1$, and make the two end-vertices of each
$A_i$ $(i=3j+2, \ j\in \mathbb{Z}^{+})$ be adjacent, to get the
second graph $G_2$. It is easy to see that $G_1$ is dynamically
3-colorable if and only if $G_2$ is dynamically 3-colorable. Third,
delete all the internal vertices in $A_i$ $(i=3j+3, \ j\in
\mathbb{Z}^{+})$ of $G_2$, and make the two end-vertices of each
$A_i$ $(i=3j+3, \ j\in \mathbb{Z}^{+})$ be adjacent, to get the
third graph $G_3$. It is easy to see that $G_2$ is dynamically
3-colorable if and only if $G_3$ is dynamically 3-colorable. Fourth,
consider the subgraphs $A_3$ in $G_3$, and there will be two kinds
of $A_3$ in $G_3$: one kind is denoted by $A_3^1$, in which the two
end-vertices of $A_3$ is adjacent (it means that the internal vertex
is contained in a triangle), the other kind is denoted by $A_3^2$,
in which the two end-vertices of $A_3$ is nonadjacent (it means that
the internal vertex is not contained in a triangle). We delete all
the internal vertices in $A_3^2$ of $G_3$, and make the two
end-vertices of each $A_3^2$ be adjacent, to get the fourth graph
$G_4$. It is easy to see that $G_3$ is dynamically 3-colorable if
and only if $G_4$ is dynamically 3-colorable. By noticing that in
$G_4$ every vertex is contained in a triangle, we have that $G_4$ is
dynamically 3-colorable if and only if $G_4$ is 3-colorable. As a
consequence, $G$ is dynamically 3-colorable if and only if $G_4$ is
3-colorable, and it is polynomially solvable whether $G_4$ is
3-colorable since $\Delta(G_4)=3$. Because we can get $G_4$ from $G$
in polynomial time, the Dy-3-Col is polynomially solvable when $G$
is in $\mathscr{C}$. \qed

For traditional colorings, the only graph $G$ with $\Delta(G)\leq3$
which is not 3-colorable is $K_4$ by Brook's theorem. By the proof
of Theorem \ref{thm4} we can easily get that there is only one class
of graphs in $\mathscr{C}$ which are not dynamically 3-colorable.
The graphs in the exceptional class, denoted by $\mathscr{E}$, can
be gotten by using a $P_i$ ($i=3j$ or $3j-1,\ j\in \mathbb{Z}^{+}$)
to replace an edge of $K_4$.

For the graphs in $\mathscr{C}$ we can determine whether they are
dynamically 3-colorable in polynomial time and we have also
characterized the exceptional graphs.

In next section, we will give a linear time algorithm to recognize
the graphs in $\mathscr{C}$ and another linear time algorithm to
determine whether the graphs in $\mathscr{C}$ are dynamically
3-colorable. At last we will give a linear time algorithm to color
the graphs by 3 colors such that the adjacency condition and the
double-adjacency condition are both satisfied.

\section{Linear time algorithms}

First, we will give a linear time algorithm to recognize the graphs
in $\mathscr{C}$. The input is a graph $G=G(V,E)$ with $|V|=n$. The
following are the main steps of the {\it recognition algorithm}.
\begin{alg}[Recognition Algorithm]\label{alg1}
\end{alg}
\begin{description}
\item[step 1.]Check if the degree of every vertex in $G$ is not more than
3. If not, return the answer that $G$ is not in $\mathscr{C}$;
otherwise, go to step 2.
\item[step 2.]Check if the graph $G$ is claw-free. If not, return the
answer that $G$ is not in $\mathscr{C}$; otherwise, go to step 3.
\item[step 3.]Check if the graph $G$ is $A_i$-free $(i=3j+1, \ j\in
\mathbb{Z}^{+})$. If not, return the answer that $G$ is not in
$\mathscr{C}$. Otherwise, return the answer that $G$ is in
$\mathscr{C}$.
\end{description}

The following are the complexity analysis of the Recognition
Algorithm: It is obvious that step 1 can be done in $O(n)$ time. If
$\Delta(G)\leq 3$, we go to step 2, otherwise $G$ is not in
$\mathscr{C}$. Since $\Delta(G)\leq3$, we just need to check the
vertices of degree 3. For every vertex of degree 3, if there is no
claw in the subgraph induced by the vertex and its neighbors, $G$ is
claw-free. So step 2 can be done in $O(n)$ time. If $G$ is
claw-free, we go to step 3, otherwise $G$ is not in $\mathscr{C}$.
In step 3, we just need to check the edges whose two incident
vertices are of degree 2 in $G$. If the paths induced by the edges
are not $P_i$ ($i=3j-1,\ j\in Z^{+}$), then $G$ is $A_i$-free
$(i=3j+1, \ j\in \mathbb{Z}^{+})$. If $G$ is $A_i$-free $(i=3j+1, \
j\in \mathbb{Z}^{+})$, then $G$ is in $\mathscr{C}$, otherwise $G$
is not in $\mathscr{C}$. Since the number of edges in $G$ is no more
than $\frac{3}{2} n$, step 3 can be done in $O(n)$ time.

Second, we give a linear time algorithm to determine if a graph in
$\mathscr{C}$ is dynamically 3-colorable. The input is a graph
$G=G(V,E)$ in $\mathscr{C}$ with $|V|=n$. The following are the main
steps of the {\it determination algorithm}.
\begin{alg}[Determination Algorithm]\label{alg2}
\end{alg}
\begin{description}
\item[step 1.]Check if there is a vertex of degree 1. If so,
return the answer that $G$ is dynamically 3-colorable; otherwise, go
to step 2.
\item[step 2.]Find the number of vertices whose degrees are 3. If the
number is not 4, return the answer that $G$ is dynamically
3-colorable; otherwise, go to step 3.
\item[step 3.]Check if the graph $G$ is in $\mathscr{E}$. If so,
return the answer that $G$ is not dynamically 3-colorable;
otherwise, return the answer that $G$ is dynamically 3-colorable.
\end{description}

The following are the complexity analysis of the Determination
Algorithm: It is easy to see that step 1 can be done in $O(n)$ time.
If there is a vertex with degree 1, the graph $G$ is dynamically
3-colorable. If there is no vertex with degree 1, we go to step 2.
Step 2 can also be done in $O(n)$ time. If the number of vertices of
degree 3 is 4, we go to step 3. In step 3, we just need to consider
the edges whose two incident vertices are of degree 2 in $G$. If
there is no such edge, $G$ is dynamically 3-colorable if and only if
$G$ is not $K_4$. If there are some such edges, we can determine if
$G$ is dynamically 3-colorable by the subgraph induced by such
edges. If there are more than one path in the subgraph, $G$ is
dynamically 3-colorable. Otherwise, $G$ is in $\mathscr{E}$, and is
not dynamically 3-colorable. Since the number of edges in $G$ is no
more than $\frac{3}{2}n$, step 3 can be done in $O(n)$ time.

Third, we will give an $O(n)$ time algorithm to color the graphs in
$\mathscr{C}$ by 3 colors such that the adjacency condition and the
double-adjacency condition are both satisfied. The input is a graph
$G=G(V,E)$ with $|V|=n$. Before we give the algorithm, we will
define a set of graphs, denoted by $\mathscr{T}$, and give some
results about the graphs in $\mathscr{T}$.

The graphs in $\mathscr{T}$ are constructed by the following two
steps:
\begin{itemize}
\item[(1)] Construct even number of vertex-disjoint triangles (3-cycles), and
the set of edges in the triangles is denoted by $E_1(G)$;
\item[(2)] For each triangle, let every vertex of the triangle be
connected by an edge to a vertex of another triangle, to construct a
3-regular graph. And the set of added edges in this step (it means
that the set of the edges are not in any of the triangles) is
denoted by $E_2(G)$.
\end{itemize}

\begin{lem}\label{lem1}
For any graph $G$ in $\mathscr{T}$, if there are $n$ triangles in
$G$, then $\alpha(G)=n$.
\end{lem}
\pf By the special structure of $G$ that every vertex is contained
in a triangle, we can see that $\chi(G)=3$ and the three color
classes have the same cardinality $n$. So, $\alpha(G)\geq n$. If
$\alpha(G)>n$, there must be a triangle that contains 2 vertices in
the maximum independent set, which is impossible, and so
$\alpha(G)=n$.\qed

\begin{lem}\label{lem2}
Let $G$ be in $\mathscr{T}$ and there are $n$ triangles in $G$. For
any maximum independent set $S$ of $G$, we have that $G\setminus S$
is bipartite.
\end{lem}
\pf By Lemma \ref{lem1}, we know that $|S|=n$, and every triangle
contains a vertex in $S$. So, in $G\setminus S$ the degrees of the
vertices are 1 and 2. Then the components of $G\setminus S$ are only
vertex-disjoint paths and cycles. Furthermore, no two edges in
$E_1(G)$ are adjacent in $G\setminus S$. Also, no two edges in
$E_2(G)$ are adjacent in $G\setminus S$. So, the cycles in
$G\setminus S$ must be alternating and have even number of edges,
i.e., there are no odd cycles in $G\setminus S$, and thus
$G\setminus S$ is bipartite. \qed

\begin{lem}\label{lem3}
For any $G$ in $\mathscr{T}$, we can find a maximum independent set
$S$ of $G$ in linear time.
\end{lem}
\pf The algorithm is given as follows:
\begin{itemize}
\item[(a)] We contract every triangle in $G$ into a vertex, and if
there are two edges which are incident to the same two end-vertices,
we can delete any one of the two edges, to get a simple graph $G'$;
\item[(b)] By Depth-First or Breadth-First algorithm, we can
find a spanning tree of a graph in $O(|V|+|E|)$ time. In our case,
it is a linear time algorithm to find the spanning tree $T'$ of
$G'$;
\item[(c)] Find a vertex $v_r$ which is adjacent to a leaf $v_l$
in $T'$ as the root of the tree. If $d(v_r)=3$, we delete $v_l$ to
ensure $d(v_r)=2$ in the new tree $T_c=T'\setminus\{v_l\}$ we will
consider later. If $d(v_r)=2$, then $T_c$=$T'$;
\item[(d)] For every edge $e_c$ in $T_c$, there are two vertices $v_s$
and $v_f$ incident to it, and $v_s$ is the child of $v_f$. By step
(a), we know that there is an $e_o$ in $G$ corresponding to $e_c$ in
$T_c$, and there is a vertex $v_o$ in the triangle corresponding to
$v_s$ which is incident to $e_o$. So, we can define an injection $f$
from $E_c$, the set of edges in $T_c$, to $V(G)$ such that
$f(e_c)=v_o$. Then we can find a vertex set $f(E_c)\subset V(G)$,
and it is easy to see that $f(E_c)$ is an independent set;
\item[(e)] Consider the three vertices in the triangle in $G$ corresponding
to $v_r$, there is just one vertex $v_r^o$ of the three which can be
added into $f(E_c)$ such that $\{v_r^o\}\cup f(E_c)$ is an
independent set. If $d(v_r)=2$ in $T'$, then let $S=\{v_r^o\}\cup
f(E_c)$, which is a maximum independent set of $G$. If $d(v_r)=3$,
there is still a triangle in $G$ corresponding to $v_l$ in $T'$ that
needs us to consider. Notice that except the vertex in the triangle
which is adjacent to $v_r^o$, any one of the other two vertices in
the triangle (assume the vertex we choose is $v_l^o$) can be added
into $\{v_r^o\}\cup f(E_c)$ such that $S=\{v_r^o,v_l^o\}\cup f(E_c)$
is a maximum independent set of $G$.
\end{itemize}

One can easily see that the algorithm described above can be done in
linear time. So we have proved the lemma. \qed

\begin{figure}[h]
\begin{center}
\psfrag{v}{$v$} \psfrag{u}{$u$}
\includegraphics[width=9cm]{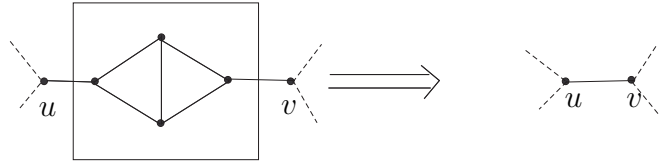}
\caption{It is the transformation in step 4 of the algorithm
\ref{alg3}. The left graph in the rectangle is the subgraph
$K_4^-$.}\label{eps2}
\end{center}
\end{figure}

Now, it is time to give the {\it dynamic 3-coloring algorithm} for a
given graph $G=G(V,E)\in \mathscr{C}$ with $|V|=n$. The following
are the main steps of the algorithm.
\begin{alg}[Dynamic 3-Coloring Algorithm]\label{alg3}
\end{alg}
\begin{description}
\item[step 1.]Delete all the pendant paths except the vertices with
degree 3, to construct a graph $G_1$, if there are some pendant
paths in $G$;
\item[step 2.]If there are some $A_i$ ($i>3$) and $A_3$ whose two
end-vertices are nonadjacent in $G_1$, we delete all the internal
vertices and make the two end-vertices of each $A_i$ be adjacent,
then we get a graph $G_2$;
\item[step 3.]If there is a vertex in $G_2$ having degree 2, then it
must be contained in a triangle. We let the two vertices which are
adjacent to the triangle be adjacent, and delete the triangle to
construct a new graph $G_2'$; And we will do the operation again if
there is still a vertex of degree 2 in $G_2'$; Similarly, we will do
the operation at most $n$ times to construct a graph $G_3$. Then
$G_3$ is 3-regular;
\item[step 4.]If there is a subgraph $K_4^-$ in $G_3$, we do a
transformation shown in Figure \ref{eps2} to get a graph $G_3'$. And
we will do the transformation again if there is a subgraph $K_4^-$
in $G_3'$. Similarly, we will do the transformation at most $n$
times to get a graph $G_4$ which does not contain the subgraph
$K_4^-$;
\item[step 5.]Now $G_4$ is in $\mathscr{T}$. By Lemma \ref{lem3}, we
can find a maximum independent set $S$ in $G_4$ in linear time;
\item[step 6.]By Lemma \ref{lem2}, we know that $G_4\setminus S$ is
bipartite, so we can color $G_4\setminus S$ by 2 colors in linear
time. We give the vertices in $S$ the third color, then we have
colored $G_4$ by 3 colors;
\item[step 7.]Color the vertices deleted before to get a dynamic
3-coloring of $G$.
\end{description}

More detailed complexity analysis about the algorithm \ref{alg3}: It
is obvious that step 1 through step 6 can be done in linear time. In
step 7, we first color the vertices in $V(G_3)\setminus V(G_4)$;
second, color the vertices in $V(G_2)\setminus V(G_3)$; third, color
the vertices in $V(G_1)\setminus V(G_2)$; forth, color the vertices
in $V(G_1)\setminus V(G_2)$; at last, we color the vertices in
$V(G)\setminus V(G_1)$. In each sub-step of step 7, we can easily
find a linear time algorithm to color the vertices such that the
adjacency condition and the double-adjacency condition are satisfied
in every $G_i$ ($i=1,\ldots,4$) and $G$. So, the algorithm
\ref{alg3} is an $O(n)$ time dynamic 3-coloring algorithm for graphs
in $\mathscr{C}$.

At last, because the dynamic 3-coloring is also a 3-coloring, the
dynamic 3-coloring algorithm is also a 3-coloring algorithm for the
graphs in $\mathscr{C}$. Furthermore, it can also become a
3-coloring algorithm for the claw-free graphs with maximum degree 3
if we modify the algorithm a little bit.

\end{document}